\documentclass[journal]{IEEEtran}

\ifCLASSINFOpdf
\else
\fi

\usepackage[switch]{lineno}

\usepackage{graphicx}
\usepackage{xcolor}
\usepackage{threeparttable}
\usepackage{multirow}
\usepackage{url} 
\usepackage{hyperref}
\usepackage{breakurl}
 \hypersetup{
    colorlinks = true,
    linkcolor = red,
    anchorcolor = blue,
    citecolor = blue,
    filecolor = blue,
    urlcolor = blue
}
\usepackage{tabularx}
\usepackage{bigstrut}
\usepackage{amsmath}
\usepackage{amsfonts}
\usepackage{amssymb}
\usepackage{wasysym}
\usepackage{adjustbox}
\usepackage{enumitem}

\renewcommand{\paragraph}[1]{\vspace{0.1cm}\noindent{\bf{#1}.}}

\begin{document}
\title{\LARGE Towards Adversarial Control Loops in Sensor Attacks: A Case Study to Control the Kinematics and Actuation of Embedded Systems}

\author{Yazhou Tu, Sara Rampazzi, Xiali Hei 
\thanks{Y. Tu, and X. Hei are with the Center for Advanced Computer Studies (CACS), University of Louisiana at Lafayette.}
\thanks{S. Rampazzi is with the Department of Computer \& Information Science \& Engineering (CISE) at the University of Florida.}
\thanks{ }
}

\maketitle

\begin{abstract}

Recent works investigated attacks on sensors by influencing analog sensor components with acoustic, light, and electromagnetic signals. Such attacks can have extensive security, reliability, and safety implications since many types of the targeted sensors are also widely used in critical process control, robotics, automation, and industrial control systems.

While existing works advanced our understanding of the physical-level risks that are hidden from a digital-domain perspective, gaps exist in how the attack can be guided to achieve system-level control in real-time, continuous processes. This paper proposes an adversarial control loop-based approach for real-time attacks on process and actuation control systems relying on sensors. We study how to utilize the system feedback extracted from physical-domain signals to guide the attacks. In the attack process, injection signals are adjusted in real time based on the extracted feedback to exert targeted influence on a victim control system that is continuously affected by the injected perturbations and applying changes to the physical environment. 
In our case study, we investigate how an external adversarial control system can be constructed over sensor-actuator systems and demonstrate the attacks with program-controlled processes to manipulate the victim system without accessing its internal statuses.

\end{abstract}

\IEEEpeerreviewmaketitle

\section{Introduction}

Cyber-physical systems rely on sensors to perceive the physical world and control actuators to apply changes to it. 
Such control mechanisms depend on sensor data that accurately depict the physical phenomena being measured. Unfortunately, analog sensors can be susceptible to spoofing via intentional physical-level interference. 

Significant efforts have been made to study how sensing systems can be susceptible to malicious acoustic, light, and electromagnetic signals \cite{giechaskiel2019taxonomy,yan2020sok}. 
However, a main barrier in applying existing attack methods to achieve continuous, real-time control is the lack of the feedback from the victim control system.
Many works \cite{trippel2017walnut,selvaraj2018electromagnetic,tu2019trick,dayanikli2020electromagnetic} rely on tuning signals by observing statuses such as the sensor measurements and timing-related signals in the victim system to adjust attack effects. However, such approaches can be limited in attacks that require real-time adversarial control. In real-time attacks, the attack signals cannot always be fully determined before the attack because the injection of time-varying analog voltage signals is continuously taking effects and inducing perturbations in the victim control system. 
Moreover, without intrusively connecting to or tampering with the victim system, its internal statuses are not necessarily available to external attackers.

\begin{figure*}
\centering
  \includegraphics[width=0.99\textwidth]{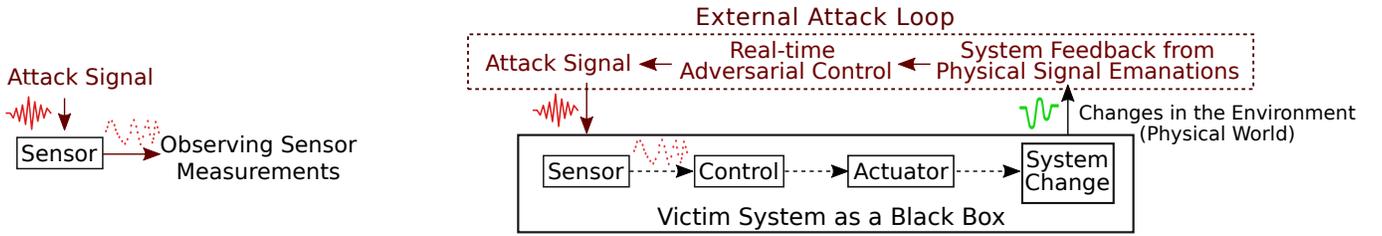}
  \caption{Transduction attacks \cite{yan2020sok} and out-of-band signal injections \cite{giechaskiel2019taxonomy} inject physically-induced voltage and current to analog circuits instead of explicit digital data to communication interfaces. 
  However, these attacks remain difficult to succeed if the systems do not have internal statuses immediately interpretable by the attacker (e.g., sensor readings printed on a screen).  
  Left: A commonly used approach is to observe statuses such as sensor outputs and tune attack signals according to the attack goal.
  Right: Our approach considers a real-time attack process that utilizes the system's physical emanations to form an adversarial loop. Leveraging this loop, we show that the attack mechanisms can be developed externally without accessing the internal data of the victim system. This approach also allows utilizing programs to automatically extract the system feedback and adjust attack signals in real-time. }\label{woot22_fig_methodology}
\end{figure*}

In this paper, we propose an adversarial control loop-based approach for real-time attacks on control systems relying on sensors. 
We study how to utilize the system feedback to guide the attacks. Specifically, by using the feedback extracted from physical-domain signals, we show that an external adversarial control loop can be constructed without requiring access to the internal statuses of the target system neither explicit sensor readings printed on a screen.

In our approach, we assume that the adversary can generate malicious physical signals to perturb the sensing system but cannot identify the exact time-varying values of induced signals in the victim system. The adversary cannot monitor the sensor module output or other internal statuses such as clock and timer signals.

We focus on real-time adversarial control processes. In this process, the input to the victim system is the perturbation caused by the attack signals. The output of the victim control system is the changes it applies to the environment. The attack system leverages the feedback to guide the attack process and continuously adjusts the perturbation to gain targeted control over the victim system.

The time series of system feedback are automatically extracted from physical-domain signals and used by programs to adjust attack signals. As illustrated in Fig. \ref{woot22_fig_methodology}, the perturbations induced by attack signals transmit to the output of the control system to actuate and induce changes to the environment. By externally observing the physical-domain actuation-related signals, we can extract features related to the system feedback.
The entire process forms a loop and allows real-time mechanisms to be developed for the attacks.

We conduct a case study to show how the adversarial control loop can be constructed on sensor-actuator systems.
We develop a real-time attack system with multiple modules running in parallel.
It captures and processes physical-domain signals related to the actuation of the victim system. By extracting features from the signals and using them as feedback, the programs in the attack system can continuously adjust the attack signals to achieve targeted process control over the victim system. The signal capturing, feature extraction and analysis, and attack signal generation are performed in multiple threads in real time while the victim system is running and being affected by the attack signals.

In our case study, we investigate attacks on kinematics-based systems that control the actuation in real time based on inertial sensor measurements. 
It requires a continuous, real-time process to manipulate the kinematics and actuation of such systems.
Without monitoring the internal hardware of the system, the conversion of attack signals in the circuits of embedded inertial sensors is subject to disturbance and is not fully predictable.
Our case study shows that external programs in the adversarial control loop can continuously extract system feedback and adjust the attack signals to manipulate the victim sensor-actuator systems.
We study the attacks on sensing and control systems including a self-balancing scooter and a prototype testing system and demonstrate the attacks with program-controlled processes to gain targeted control over the victim system without monitoring its internal hardware and sensor statuses. 
Existing works studied physical-level signal injections on sensing and actuation systems \cite{selvaraj2018electromagnetic,son2015rocking,trippel2017walnut,tu2018injected}. However, how to leverage the system feedback automatically extracted from physical-domain signals to guide real-time, continuous attack processes have not been systematically investigated.
Furthermore, our paper studies a general approach to overcome the inherent limitations of analog sensor attacks to control continuous processes by leveraging an external attack loop (Fig. \ref{woot22_fig_methodology}).
This approach allows automatic mechanisms and programs to be developed to extract the system feedback and control the attack process in real time.

In summary, this paper makes the following contributions.

\begin{itemize}

\item We study how to utilize the system feedback automatically extracted from physical-domain signals to guide real-time attacks on sensors.

\item 
We propose an adversarial control loop-based approach for attacks on process and actuation control systems relying on sensors. Our approach can overcome the uncertainty issue that limits analog sensor signal injections from achieving system-level adversarial control in continuous attack processes.

\item We conduct a case study to show how our method applies to real-world systems. We show that it is possible to construct an external adversarial control loop in sensor attacks without requiring access to the victim system's internal statuses. 
Our experimental study demonstrates real-time adversarial process control over sensor-actuator systems by utilizing program-controlled mechanisms in the external adversarial control loop.

\end{itemize}

\section{Attack Characterization and Threat Model}

In this section, we characterize real-time attacks on sensor-based control systems. We focus on attacks where adversaries need to adjust the injected signals in real time to control the victim system. 

We also specify typical types of attack settings based on different assumptions about the adversaries' capability to access the internal data and modules of the victim system.

\paragraph{Scope} 
In our analysis, we consider systems that control physical properties and apply changes to the environment in a continuous process. These systems are widely used in continuous process control, robotics, and automated systems. For instance, robotic systems~\cite{klemm2019ascento,bostonhandle_01,bostonhandle_02} continuously measure and adjust the system kinematics while keeping balance or performing specific tasks. In industrial control systems and medical devices, properties such as the temperature and gas/fluid pressure often need to be continuously monitored and controlled~\cite{thermocouple,antonucci2009infant,pressure_thomas,hashemian2011line}.

To manipulate such processes through the injection of malicious signals, an attack needs to constantly exert influence on the system to apply changes to the physical environment in a real-time controlled manner. 
We investigate how to leverage the system feedback to guide this real-time attack process.

Finally, adversaries could also use previously determined signals (e.g., recorded voice) to trigger system reactions, such as the injection of malicious voice commands in voice assistants \cite{zhang2017dolphinattack,sugawara2020light}. 
Such systems that perform an instant action based on a triggered command or event will not be the focus of this work.

\paragraph{Specification of Different Attack Settings}
Based on different assumptions about the adversaries' capability to access the internal of the victim sensing and control system, we characterize the attack settings and discuss the potential issue of analog sensor attacks that do not have access to the internal statuses of the victim system.

\textit{1) White-box attack setting}. In white-box sensor attacks, the adversary has full access to the internal statuses of the victim system. Such statuses can include the sensor module outputs, clock signals, control algorithms, and parameters. For example, the adversary can connect to the modules being tested with an Arduino or oscilloscope. The adversary can adjust and fine-tune the attack signals by directly observing the sensor outputs.

\textit{2) Gray-box attack setting}. While white-box attacks serve as great initial steps to discover vulnerabilities and understand the physical attack mechanisms, they can suffer from the uncertainty in analog signal injections when the internal of the victim system is not fully accessible.
Gray-box attacks can be used to enable the transition from the white-box to a black-box setting. 
In gray-box attack setting, the adversaries could partially know the parameters, models or monitor the readings shown on the external screen (if present) of the victim system without intrusively connecting to it.

\textit{3) Black-box attack setting}. In this setting, the adversary can still know the general working principles and observe the external behaviors of the target system. However, the internal statuses of the victim system are considered as in a black box. Black-box attacks do not require access to the internal statuses of the victim system.
Moreover, the adversary cannot connect to digital monitoring or controlling interfaces of the victim system.

Because the principle of sensor signal injection attacks is to inject analog signals instead of explicit data or control commands via digital communication interfaces, a main problem in analog sensor signal injection attacks is that the time-varying values (e.g., amounts, phases) of the injected signals are uncertain. 
For sensor attacks, the attack signals often need to be remotely transmitted to the victim system, transduced and non-linearly converted in the sensor circuits. Without relying on the white-box setting, it can be challenging to accommodate the uncertainty in real-time attacks or fully predict the phases and amounts of real-time statuses in the victim system. This paper studies real-time attacks in a black-box setting. We explain the detailed settings and assumptions as follows.

\paragraph{Threat Model}
We assume that the adversary cannot tamper with or access the internal firmware/hardware of the victim system. The adversary also cannot directly modify the actual physical property (such as by directly heating the environment with a heater to increase the actual temperature). 

We study a continuous, real-time attack process to implicitly control sensing-based control systems. In this process, the attacks manipulate the victim system in real time to apply changes to continuous physical properties in a targeted manner.
The attack approaches are different from explicit channels that directly send control commands to the system with digital-domain data communication.

The adversary cannot access the internal sensor measurements, sampling intervals, clock and timer signals of the victim system. The attack mechanisms are developed externally without connecting to the internal hardware or digital communication and control interfaces of the system.

The proposed attacks can allow adversaries to gain the capability of victim systems to implicitly control continuous physical properties in a relative manner without accessing the system's internal hardware. Depending on the application, the attack consequences can range from malicious control of embedded systems to potential real-world safety issues in critical process control systems.

Finally, the attacks are always ``online'' attacks instead of ``offline''. Offline signal processing and analysis methods could be used for post-attack analysis but cannot guide the attack process. This is because of the time-varying nature of injected signals and other statuses in continuous process control systems. Therefore, the feedback extraction and attack signal adjusting are performed in real time while the victim system is continuously being affected by the perturbations and the next statuses are not yet determined.

\section{Methodology and Attack Loop Mechanisms} \label{woot2022_sec03_methodology}

In this section, we introduce the structure of the proposed methodology and describe the real-time mechanisms to form an external adversarial loop over a sensor-actuator system. 

\subsection{Adversarial Control Loop Structure}

\begin{figure}
\centering
  \includegraphics[width=0.99\linewidth]{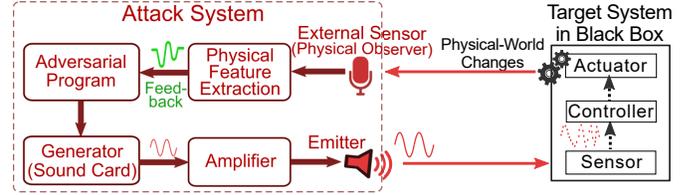}
  \caption{An illustration of the structure and basic modules of the proposed method. The input to the victim system is the perturbation caused by the attack signals. The output of the victim control system is the changes it applies to the environment (e.g., the physical world).
The system feedback is automatically captured and used by the attack system.
  }\label{fig_woot2022_modules}
\end{figure}

Fig. \ref{fig_woot2022_modules} illustrates the structure and basic modules in our approach. We model the target system as a black-box sensor-actuator system that controls its actuation based on real-time sensor measurements. 
The attack system is external to the victim system and can only affect the target system by emitting physical-domain signals.

The main modules in the attack system include the physical observer, the feature analyzer, the adversarial control program, and the attack signal generator.

The real-time feedback is provided by the physical observer and the feature analyzer modules.
The physical observer of the attack system consists of sensor modules that measure physical signals related to environmental changes. The data is streamed to the feature analyzer, which extracts the system feedback by analyzing physical signal features. 

The adversarial program on a computer or an embedded device will adjust the attack signals based on the extracted feedback. The program then generates the attack data and writes the real-time attack signal streams to a signal generator hardware. (e.g., a sound card).
Additionally, an amplifier is used to drive the signal emitting device. Depending on the type of attack signals, a transducer or antenna can be used to emit the attack signals.

During the attack process, the modules will not pause and wait for other modules. To form an adversarial control loop, the different modules are performing in parallel unless the attack is stopped or paused. We implement the system's software with multi-threaded programs that handle the inputs and outputs of each module in real-time. Later sections will introduce more details about the attack system.

\subsection{Adversarial Control Loop Formalization}

The real-time control is enabled by a loop that can inject signals to perturb the victim system and extract the system feedback from physical-domain signals. We characterize the general mechanisms including real-time analysis and control, injection of perturbations, and system feedback extraction. These mechanisms are functioning in parallel to form the real-time attack loop.

\subsubsection{\textbf{Real-Time Analysis and Control}} 
To understand how the perturbation injection can be guided to control the victim system by leveraging the extracted system feedback from physical domain signals, we provide the following analysis and formalization. 

Assuming the perturbation injected into the input of the victim system is $p(t)$, the perturbed system input $X'(t)$ is

\begin{equation}\label{woot22_formula_input}
\begin{small}
\begin{array}{l}
X'(t)=X(t)+p(t),
\end{array}
\end{small}
\end{equation}

where $X(t)$ is the original status without the perturbation. The actuation output of the system under perturbation is $Y(t) = G(X'(t))$, where $G$ is the transfer function of the victim control system. For control systems that rely on the sensor measurements to perform real-time process control, the injected perturbations $p(t)$ will be transmitted to the output of the system via

\begin{equation}\label{woot22_formula_transfer}
\begin{small}
\begin{array}{l}
Y(t)=G(X(t)+p(t)). 
\end{array}
\end{small}
\end{equation}

When the goal is to control a real-time process instead of randomly interfering with it, the time-varying effect of the perturbation also has to be considered.
By extracting and analyzing the time series $y[n]$ of the system feedback $y(t)$ from physical-domain signals related to $Y(t)$, the injection of perturbations can be guided and adjusted in real-time to gain targeted control over the system.

\subsubsection{\textbf{Injection of Perturbations}} The injection of perturbation $p(t)$ is conducted by influencing  analog sensor components with physical-level signals. The exact values of the injected signals as well as other internal statuses of the victim system are not accessible for external attackers. Such injections do not directly control the behavior of the victim control system, but induce voltage signals in the sensor circuit to implicitly perturb the control system based on sensors.

\subsubsection{\textbf{System Feedback Extraction}}\label{woot22_subsubsec_feedback}
There can be many ways to retrieve the time series of the system feedback $y(t)$ without accessing its internal statuses. For instance, cameras can be used to detect and measure the movements of the actuator; infrared sensors can remotely monitor the temperature of the target environment; laser-based devices can be utilized to measure the distance and speed of the target system.

The exact injection and feedback extraction mechanisms can differ depending on the attack type, target sensors and the process or actuation control systems (Section \ref{sec_general_mechanisms}). Section \ref{sec_specify_mechanisms} will discuss the specific mechanisms in our case study.

\subsection{Model Generalization}\label{sec_general_mechanisms}

The proposed adversarial control loop attacks can be applied to different kinds of process control and actuation systems based on sensors.  We discuss the system property as follows.

\textbf{System Properties.} In continuous process and actuation control systems, the system controls a continuous physical property such as temperature, air pressure, pH value, heading angle. Typical process sensors include thermistors, thermocouples, pressure sensors, pH sensors, etc. Additionally, inertial measurement units are used in robotic systems to control the system kinematics. The control is usually also a continuous process. For instance, when a robotic system measures the heading angle and actuate the system to maintain a balanced position. Telepresence control systems or virtual/augmented reality (VR/AR) systems actuate an object in a remote or simulated environment. The heading angle or the kinematics properties of the systems are also continuous properties.
the temperature, pressure, pH value of a physical environment or the speeds of motors and heading angles of actuation systems. Such properties are continuous because they can not be changed instantly but requires a process for the system to apply changes to the physical environment.

Process actuators include motors in actuation systems, heaters in temperature control systems, peristaltic motors in pH control systems, valves, pumps, etc.
Their actuation will result in a change in the process variable. Furthermore, the operation of the mechanical and electrical parts will generate changes in environmental signals. An external attacker can measure the induced changes in the signals of the physical environment. Examples of such signals include but are not limited to heat, sound, electromagnetic, motion, visual signals, images, and electric signals.

In real-time process and actuation control, $Y(t)$ is the output of the control system and is directly related to the actuator status. For example, $Y(t)$ can be the acceleration or speed of the motor. Additionally, we use $Z(t)$ to represent the process variable of the control system. Depending on the application, $Z(t)$ can be equal to $Y(t)$ or be affected by $Y(t)$. For instance, when the control goal is to reach a certain motor rotation speed, we have $Z(t)=Y(t)$. In many other scenarios, the process variable $Z(t)$ is not the statuses of the actuator. $Z(t)$ will be affected by $Y(t)$ but will not be equivalent to it. For example, when the goal is to control the pH value, the actuator status $Y(t)$ of the peristaltic motor or other flow controlling actuators will determine the volume of acid or alkaline solutions delivered to the environment in a certain amount of time. This further determines how the actual pH value $Z(t)$ will be changed. In air pressure control systems, the actuator status $Y(t)$ will affect how quickly the air is pumped into the system or exiting, which determines the actual pressure of the system. Usually, $Y(t)$ is correlated to the change of $Z(t)$.

The adversarial control loop can apply to different kinds of systems with the following stages. First, depending on the control system, physical signals in the environment can be extracted by adversaries. In scenarios that the physical signals are directly related to the actuator, the extracted feedback will be denoted as $y(t)$, which is directly correlated to the control system output $Y(t)$. This applies for kinematics-based control systems and many other process control systems. For instance, the actuation of pumps in pH control systems can generate acoustic or electromagnetic signals. Additionally, for temperature control systems, the feedback can be extracted with an infrared sensor (e.g., handheld thermo gun). The extracted time series of $z(t)$ will be directly related to the process variable being controlled by the victim system. Usually, actuation of control systems will generate environmental changes that can be measured by acoustic, electromagnetic, current, heat, visual, or vibration signals, etc.

Second, the adversaries will inject physical-domain signals to induce malicious current or voltage signals in the analog sensor of a victim system. For example, adversaries can inject out-of-band acoustic signals to inertial sensors to perturb their readings \cite{son2015rocking,trippel2017walnut,wang2017sonic,tu2018injected}. Adversaries can affect the output of pH, temperature, and pressure sensors by EMI attacks\cite{tu2019trick}. 

Finally, by real-time processing and analysis of the extracted feedback, the injected perturbation $p(t)$ can be adjusted with algorithms to control the victim system. The methodology provides a systematic approach to control the victim embedded system without connecting to the internal statuses or modules. Further, different from relying on empirical observation and manual signal tuning, it provides a formalized, universal framework with program-controlled processes for different instances of attacks.



\subsection{Adversarial Control Loop in the Case Study}\label{sec_specify_mechanisms}

In this section we show how our method applies on real-world attack scenarios. We investigate adversarial control loop-based attacks to gain targeted control over kinematics-based control systems (e.g., self-balancing scooter). The attack system perturbs the sensor data with resonant acoustic signals and adjusts the attacks signals based on the system feedback. We specify the mechanisms utilized in our case study as follows.

\subsubsection{System Feedback Extraction}\label{woot22_subsubsec_feedback}
We utilize a microphone to collect the sound of the actuator. Through experiments and analysis, we find that the sound emanation of motors can be leveraged to extract the system feedback to guide a real-time attack process. Specifically, we provide the following formalization.

Assuming the output mechanical power of a motor is $P$, measured in watts (W), the rotational speed is $rpm$ in revolutions per minute. The work done per revolution in Joule is $Work = Force \cdot Distance = \frac{\tau}{Radius}\cdot 2\pi Radius = \tau\cdot 2\pi$, where $\tau$ is the torque of the system. The output power of the motor $P$ is related to its speed $rpm$ in the following formula:

\begin{equation}\label{ch01_imu_eq4_2}
\begin{small}
\begin{array}{l}
P = \tau \cdot rpm \cdot 2\pi / 60,
\end{array}
\end{small}
\end{equation}

where $\tau$ is the torque of the system, $P$ and $rpm$ are directly correlated. 
Since the energy of acoustic emanations $P_s$ is also related to the power of the motor (e.g., by friction, vibration, coil noise), for simplicity, we assume that $P_s$ is related to $P$ by $P=\alpha P_s $, where $\alpha$ is a constant value to describe the ratio between the acoustic emanation power and the motor power. We have $\alpha P_s  = \tau \cdot rpm \cdot 2\pi / 60$. The formula is not used to accurately determine the acoustic energy or the speed but can serve as an empirical tool in relative estimations. 
Our experimental results in later sections also show that the extracted time series $y[n]$ from acoustic emanations can be highly correlated with the actual motor speed and movement patterns. Therefore, we can utilize the extracted and processed features to guide real-time attacks.

This methodology can be adapted to using multiple feedback signals.
Actuation can result in various physical-level signals in the environment. 
Depending on the type of the target system and actuators, the feedback can be extracted from acoustic signals, heat, electromagnetic signals, motions, or visual signals. 
Generally, such signals can be measured by different kinds of sensors such as cameras, radars, microphones, infrared thermometer, laser-based distance and speed sensors, etc. The system feedback can also be measured using a combination of different kinds of sensors.

Both power and frequency related signals can be useful feedback to guide the attacks.
We will explain the details of how to automatically extract and utilize these features in Sections \ref{sec_design_eva} and \ref{sec_prototype_experiments}.

\subsubsection{Injection of Perturbations}

The measurements of micro-electromechanical systems (MEMS) inertial sensors can be interfered with by acoustic signals due to their susceptibility to acoustic resonance \cite{dean2010characterization,son2015rocking}. The high-frequency acoustic signals injected into the sensors can be converted to low-frequency in-band signals by aliasing \cite{trippel2017walnut}. Further, this conversion process is not perfect and can be subject to the disturbance caused by the real-time drifts of sampling intervals in embedded systems \cite{tu2018injected}. 

Assuming the attack signal $m(t)$ is:

\begin{equation}\label{ch01_imu_eq4_1}
\begin{array}{l}
m(t) = A \cdot sin(2\pi f(t)t+\phi_m),
\end{array}
\end{equation}

where $A$ and $f(t)$ are the amplitude and frequency of the attack signal. $\phi_m$ is the initial phase. After being transduced in the sensor, the injected signal becomes $V(t)$.

\begin{equation}\label{ch01_imu_eq4_1}
\begin{array}{l}
V(t) = A_0\cdot sin(2\pi f(t)t+\phi_0),
\end{array}
\end{equation}

where $A_0$ and $\phi_0$ are the amplitude and initial phase of $V(t)$. While attackers have full knowledge about the signal $m(t)$, they may not have full knowledge about $V(t)$. For instance, the initial phase $\phi_0$ is not certain for attackers after signal transmission and conversion. 

Moreover, the frequency of the injected signal after aliasing is also not fully deterministic. Assuming $\Delta T[i] = \delta[i] + \frac{1}{F_{S0}}$ are the sampling intervals. $F_{S0}$ is the ideal sample rate. $\delta[i]$ is the drift in the sampling interval. The exact value of $\delta[i]$ can be affected by imperfect clock signals or other kinds of software delays or interrupts in the victim system. Therefore, it is unlikely for adversaries to fully predict the values of the exact sampling intervals in real time. $t_0=0, t_1=\Delta T[1],..., t_i=\sum\nolimits_{j=1}^{i}\Delta T[j],...,$ are sampling times. The digitized signal $V[i]$ will be

\begin{equation}\label{woot22_eq_digitizedrifts}
\begin{small}
\begin{array}{l}
V[i] = A_0\cdot sin(2\pi \epsilon(t_i) t_i + 2\pi nF_{S0} (\sum\nolimits_{j=1}^{i}\delta[j])+\phi_0).  
\end{array}
\end{small}
\end{equation}

Ideally, the digitized signal would have a frequency of $\epsilon(t)$, assuming $f(t) = nF_{S0} + \epsilon(t) (-\frac{1}{2}F_S < \epsilon \le \frac{1}{2}F_S,n\in \mathbb{Z}^+ )$. However, due to the drifts in sampling intervals, the frequency of the digital signal is also subject to disturbance. We can observe a non-constant term $2\pi nF_{S0} (\sum\nolimits_{j=1}^{i}\delta[j])$ in the signal. This is an accumulated term amplified by $n$. It changes over time and brings real-time disturbance to the converted signal. 

Because of drifts in sampling intervals, the induced sensor signals will be oscillating signals. The value of the signal also fluctuates as the phase of the signal changes in real time. Such fluctuating signals will be interpreted as noises by kinematics-controlled systems relying on inertial sensors.
We model the injected signals as perturbations that influence the inputs on the victim system. Further, the victim control system affected by the perturbation will make changes in the environment in real time. We then leverage the system feedback to accommodate the uncertainty and adjust the perturbation to achieve targeted adversarial control in real-time attacks.

\subsubsection{Real-Time Analysis and Control}

In our case study, the control systems affected by the injection will exhibit an oscillating pattern in actuation. This is because the injected signal is oscillating and the real-time changes in sensor data are transmitted to the output of the victim control system via the transfer function (Eq. \ref{woot22_formula_transfer}). 
The feedback $y(t)$ that describes $Y(t)$ can provide necessary information for controlling the attack process. For instance, the relative time-varying power and frequency of the actuation under the perturbation can be analyzed and used by programs to gain targeted control.

Based on the feedback time series $y[n]$, we can develop automatic mechanisms that selectively apply existing signal injection techniques to adjust $p(t)$ in a real-time, program-controlled manner. For example, 
Tu et al. proposed two kinds of attacks to manipulate systems based on inertial measurement units (IMUs) \cite{tu2018injected}. The Switching attack intentionally manipulates the oscillating digitized signal by repetitively switching the out-of-band attack signal frequency $f(t)$ \cite{tu2018injected}. The Side-Swing attack adjusts the attack signal amplitude $A$ within each oscillation cycle of the induced perturbation to manipulate the accumulated effect \cite{tu2018injected}. Our prototype attack system will selectively apply these signal injection techniques in real time to adjust $p(t)$ in Sections \ref{sec_design_eva} and \ref{sec_prototype_experiments}. 
Instead of focusing on designing specific analog sensor signal injection techniques, we investigate a general approach and study utilizing automatic mechanisms guided by extracted feedback time series in the context of an external adversarial control loop.

\section{System Design and Attack Validation}\label{sec_design_eva}

In this section, we design and implement the prototype real-time attack system. We discuss the general procedure to achieve the external adversarial control loop on sensor-actuator systems and validate the methodology on a self-balancing scooter in black-box settings.

Without the attack loop, adversaries need to manually adjust the attack signals while observing the internal statuses \cite{trippel2017walnut,selvaraj2018electromagnetic} or the victim system behavior \cite{tu2018injected}. The differences between our method and existing attacks are the automatic extraction of system feedback time series from physical-domain signals and real-time mechanisms that continuously adjust the attack signals in the context of an external adversarial control loop. Our continuous manipulation processes are facilitated and adjusted with program-controlled mechanisms instead of relying on the reaction speed and concentration of human attackers. Our procedure non-intrusively connects the physical sensor and actuator parts of the victim system to the real-time mechanisms and programs in the external attack loop.

\begin{figure}
\centering
  \includegraphics[width=0.88\linewidth]{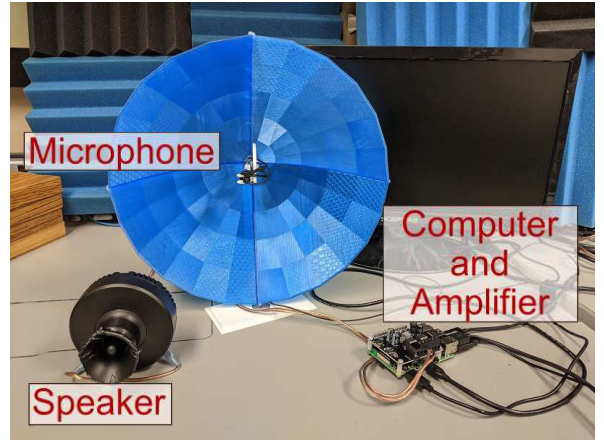}
  \caption{The hardware components of the testing system. We 3D print a parabolic reflector for a cheap lavalier microphone. The computer can be a desktop/laptop or an embedded device such as a Raspberry Pi.}\label{fig_woot22_hardware_setting}
\end{figure}

\subsection{Settings}\leavevmode

\noindent

In the attack system, the programs on the computer extract the system feedback from the physical-domain information automatically and utilize the feedback to adjust attack signals in real time. The attack signals are fed to an external digital-analog-converter (sound card) that connects to an amplifier. 
We use a high-output-resolution ($\geq$96kHz) sound card and a tweeter speaker to generate the attack signals. 
Additionally, a microphone captures the physical acoustic signals emanated by the target system.  
We 3D print a parabolic reflector \cite{parabolic_mic} for the microphone (Fig. \ref{fig_woot22_hardware_setting}) to pick up signals in a target area. The reflector can be produced with a low-end 3D printer.

We implement the software modules of the multi-threaded attack system in Python and run them on a Raspberry Pi to perform signal processing, feedback extraction, analysis, attack signal adjusting in real time. 
We implement the software signal injector module in C. It allows attack signals to be generated and adjusted in real time while writing data to the sound card without introducing extra latency. Such latency can lead to glitches and discontinuity in the attack signals. The module directly uses the low-level sound APIs (Linux ALSA \cite{alsawiki}) to control the sound card.  

Additionally, the actual motor speed of the victim system is measured with a hall effect switch. It is used to provide  quantitative analysis, and its data is not used for attacks.

\begin{figure}
\centering
  \includegraphics[width=0.98\linewidth]{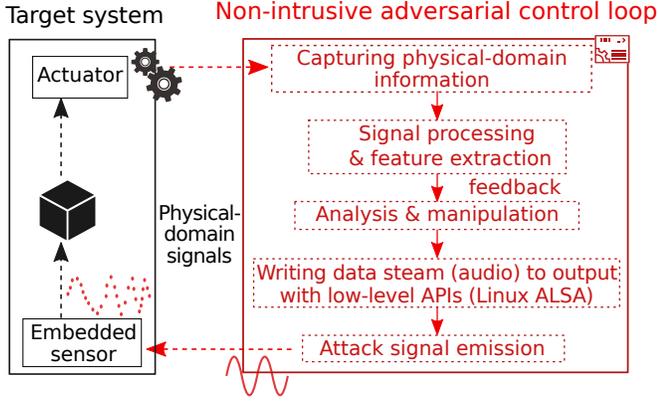}
  \caption{An illustration of the procedure and main functionalities of the adversarial control loop-based approach. The attack system is multi-threaded and performs its functionalities in parallel. }\label{woot21_fig_procedure}
\end{figure}

\begin{figure*}
\centering
  \includegraphics[width=0.99\textwidth]{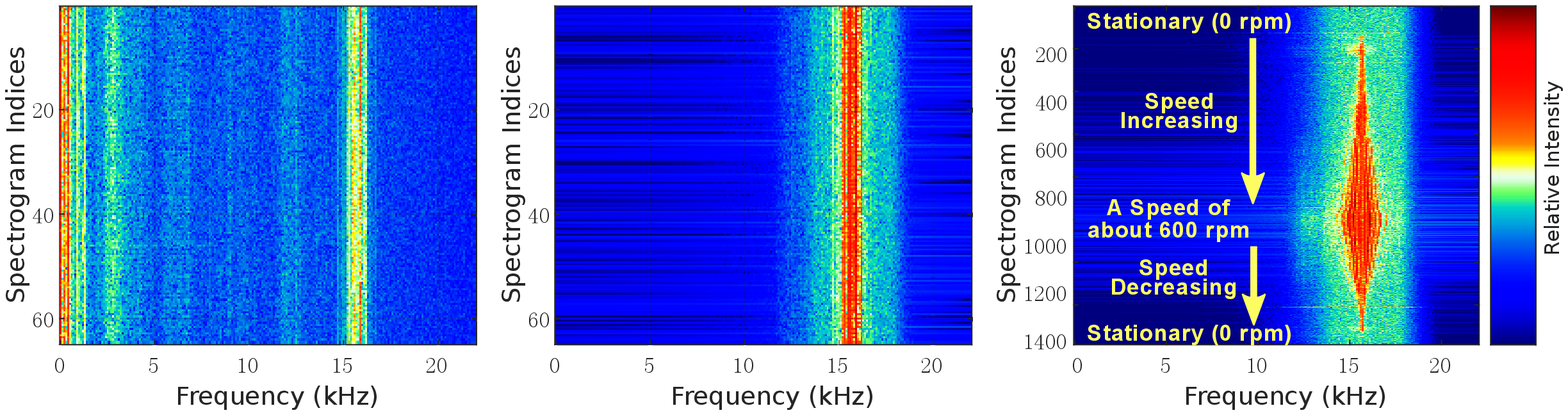}
  \vspace{-0.5mm}
  \caption{Left: The spectrogram of the captured acoustic signals (unfiltered) in a segment of 15 seconds while the motor is rotating with a speed of about 300 rpm. Middle: The spectrogram of signals under the same condition but with Butterworth band-pass filtering. Right: The spectrogram of the filtered acoustic signals when the motor speed is changing. We gradually change the motor speed by manually adjusting the actual heading angle of the self-balancing scooter. }\label{fig_woot22_spec_magawheel}
\end{figure*}

\subsection{Validation}\leavevmode \label{sec_black_method}

We characterize the general procedure to construct the adversarial control loop (Fig. \ref{woot21_fig_procedure}). The tasks in different modules are performed in parallel with multiple threads in real-time attacks to affect the victim system continuously.
We explain the main steps and functionalities as follows.

\paragraph{Capturing physical-domain signals}
In our case study, we capture acoustic signals emanated by the actuator of the target system. 
It is not required to use this specific method to observe and capture the physical information.
A combination of other methods leveraging cameras, radars, or laser-based sensors can also be utilized in the future.
We utilize the parabolic microphone as a directional receiver of signals emitted by the target. Other microphones can be used as well. However, using a parabolic microphone is more effective in noisy and long-distance scenarios.

\paragraph{Signal Processing and Feedback Extraction}
The program first applies real-time fast Fourier transform (FFT) to get frequency-domain features with a window size of 4,096 samples. If this window size is too small, the spectrum analysis can be less effective. However, if it is too large, the system reaction time will be slow. We select 4,096 to allow an update time of less than 0.1 seconds ($0.093s$ with a recording sample rate of 44.1 kHz).

Fig.\ref{fig_woot22_spec_magawheel} shows the frequency analysis results of acoustic signals recorded while the motor of the Megawheels self-balancing scooter is moving.
We can identify that the range of sound frequencies of the actuator is from $14,600$ Hz ($F_l$) to $16,900$ Hz ($F_h$).
By utilizing a Butterworth bandpass filter, we can remove noises in other ranges and observe the relationship between the motor speed and the frequency-domain features more easily.

Since the strength of frequency components in the identified range $[F_l, F_h]$ is highly related to the speed of the motor (Fig.\ref{fig_woot22_spec_magawheel} right), we use the sum of the magnitude of all frequency components in this range as the system feedback. The attack program extracts the feedback $y(t)$ data stream in time series 

\begin{equation}\label{ch01_imu_eq4_2}
\begin{small}
\begin{array}{l}
y[n] = \sum_{f=F_l}^{F_h} R(f,n\cdot T_c), 
\end{array}
\end{small}
\end{equation}

where $T_c\approx0.093$ is the duration of each chunk of signals being processed, and $R(f, t)$ is the magnitude of the frequency component $f$ at a time window $[t-T_c, t]$. Given a specific time $t$, only the time series of $y[1],...,y[n]$ with $n\cdot T_c < t \ (n \in  \mathbb{Z}^+) $ are available during the attack process.

Additionally, we observe that the actuator can generate a part of electrically and mechanically induced acoustic noises that are less relevant to its speed. Such noises can lead to small spikes in the time series of $y[n]$.
We can effectively mitigate this effect with a simple weighted moving average filter. In detail, we get $ y_0[n] = \sum_{i=0}^{k}w_i y[n-i]$. A small window size (e.g., $k=4$) works well to reduce the noise while maintaining the sensitivity. We have $ y_0[n] = \frac{6}{18} y[n] + \frac{5}{18} y[n-1] + \frac{4}{18} y[n-2] + \frac{3}{18} y[n-3]$. A larger value of $k$ could increase the signal smoothness but would also make the feature less sensitive to reflect the actual changes. During the attack, the program automatically processes the signals and computes the feedback in real time.

\paragraph{Analysis and Manipulation} 
Suppose that the injection induces perturbations in the system input by $X'(t)=X(t)+p(t)$ (Eq. \ref{woot22_formula_input}), the perturbations will then be transferred to the output of the system via $Y(t)=G(X(t)+p(t))$ (Eq. \ref{woot22_formula_transfer}).
Therefore, the functionality of this step is to selectively adjust the attack signals to change input perturbations $p(t)$ in real time based on the extracted feedback $y(t)$ in order to achieve targeted adversarial control over the actuation of the victim control system.

We develop automatic real-time mechanism to gain control over the sensor-actuator systems. To adjust the injected perturbations in real time, we design the following dynamic threshold setting method and utilize the signal injection technique of the Switching attack \cite{tu2018injected} that injects phase offsets by repetitively switching the attack frequency.

\noindent
\textit{Dynamic threshold setting.} Different from manual attacks, the attack system rely on the extracted feedback to adjust the attack signals. We develop a dynamic mechanism for the attack system to set a threshold and perform frequency switching operations automatically during the manipulation stage.

The attack system monitors the value of the recent peak ($K$) in the time series of the feedback $y_0[n]$ and sets the threshold as $T_h = \alpha K - \beta K^2$. It switches the frequency when the value of $y_0[n]$ drops down and crosses this threshold.
Since we want the program to switch the attack frequency when the signal is at a higher level for efficiency, we set $\alpha$ as a value close to $1.0$ (such as $0.95$). Selecting a larger threshold can also help compensate for the slight delay ($\delta$) in the control system and the signal streaming in the adversarial control loop. We set $\beta$ as $0$ or a minimal value to adjust the threshold when $y_0[n]$ reaches a large value (such as when the motor is rotating with the maximum speed). 

Additionally, the drifts of sample rates can cause disturbance in the signals and accumulate over time (Eq. \ref{woot22_eq_digitizedrifts}).  To mitigate this effect, the attack system records two intervals between the last frequency switching operations and uses their ratio to adjust the center frequency.
Further, Switching attacks with a smaller difference (step size) between the attack frequencies result in a smaller frequency of the induced movements and can be used to induce a larger accumulative effect in the victim system within a certain amount of time. Therefore, the manipulation process starts with a larger step size and updates it by $step' = \gamma \cdot step, 0.8 < \gamma < 1.0$ in reach round of adaptation until a minimum step size is reached.
In our experiments, we set the initial step size as 1.5 Hz and the program would decrease it until it reaches a minimum value of 0.85 Hz during the automatic manipulation process.

\begin{figure}
\centering
  \includegraphics[width=0.98\linewidth]{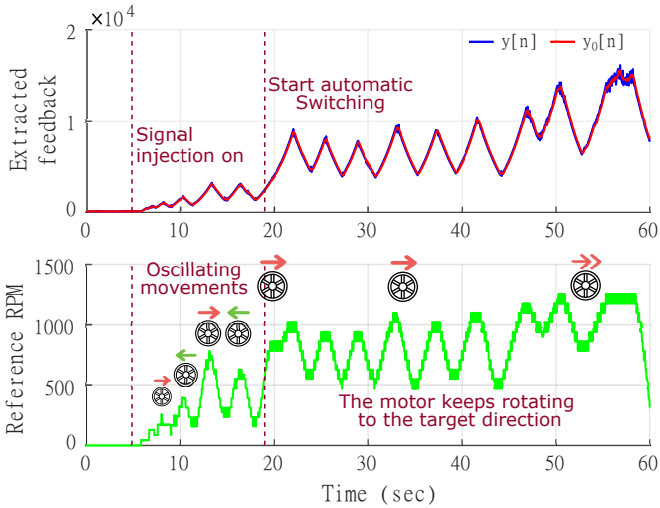}
  \vspace{-0.8mm}
  \caption{Real-time attack on the sensor-actuator system of a self-balancing scooter. Before the signal injection, the scooter motor is stationary (0-5 s). The signal injection induces oscillating movements in the system (5-19 s). After the automatic Switching process starts (at 19 s), the attack system controls the manipulation process based on the extracted feedback in real time. The reference RPM is measured with a hall effect switch to show the actual speed of the motor and is not used by the attack system.
}\label{auto_blackbox_hoverboard}
\end{figure}

\paragraph{Validation}
We demonstrate the proof-of-concept attack on the Megawheels self-balancing scooter in a black-box setting. The speakers are placed in proximity (3-5 cm) to the scooter. This range can be extended by using a higher volume and directivity horns \cite{tu2018injected,horn_2}.
Under resonant acoustic signal injections, the scooter presents an oscillating movement pattern that alternatively turns the wheel in different directions. This is because that the oscillating signals induced in the sensor are perturbing the system. Such signals can induce shaking and oscillating movements in the victim system but do not present targeted control.

Fig.\ref{auto_blackbox_hoverboard} illustrates this process. When the resonant acoustic signals turn on, the injected perturbations induce
oscillating movements in the motor of the scooter.
After the automatic Switching process starts, the attack system controls the attack process in real-time based on the time series of $y_0[n]$.
During this manipulation process, the attack system automatically extracts the feedback and adjusts the attack signals.\footnote{Anonymized demos of the proof-of-concept attacks are available at \url{https://www.youtube.com/playlist?list=PL_l1Kb3yQ2-ZllwC31CqIG5dNzJTXObF_}.}  By utilizing the dynamic threshold and the feedback data, the attack system performs frequency switching operations in real-time to induce targeted control effects. In the automatic Switching process, the attack system can turn the scooter wheel into a target direction and manipulate its speed. As shown in Fig. \ref{auto_blackbox_hoverboard}, from 19 s, the motor keeps rotating to the target direction and reaches a high speed by the end of the attack. The induced speed can be further adjusted in automatic Switching processes by using a different amplitude.
The manipulation is a continuous process because the victim control system is not controlled with an instant value or event, but requires continuous control to selectively perturb the time-varying statuses and apply changes to the environment.

\section{Experiments with a Prototype Testing System}\label{sec_prototype_experiments}

In this section, we develop a prototype testing system to study attacks on kinematics-controlled sensor-actuator systems. We study the attacks with  program-controlled processes in the adversarial control loop and discuss our observations.

\subsection{Prototype Testing System}
\begin{figure}
\centering
  \includegraphics[width=0.9\linewidth]{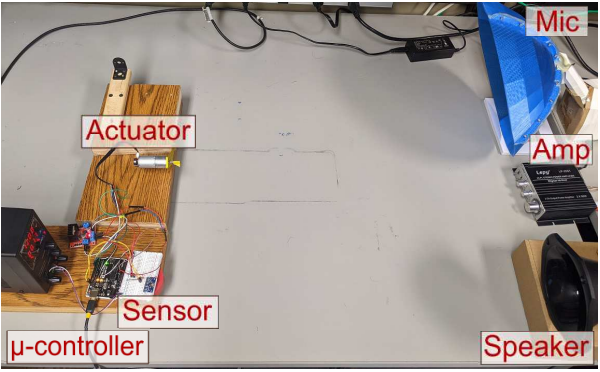}
  \caption{Our prototype testing system. In the victim system (on the left), a micro-controller controls the actuator in real-time based on the sensor measurements. The attack system (on the right) manipulates the victim system in real-time via the external adversarial control loop. The two systems are not interconnected with any communication or control interfaces.}\label{fig_woot22_prototypetesting}
\end{figure}

Fig. \ref{fig_woot22_prototypetesting} shows the testing system including a victim system and an attack system.

In the victim system, the micro-controller (Arduino Uno) controls the motor in real time based on kinematics measurements.
It is a real-time sensing and control system that sets the motor speed based on the heading angle measured with the gyroscope.
Specifically, it measures the heading angle of the system by using the gyroscope sensor of the inertial measurement unit (MPU-9250). The actuator is a geared DC motor equipped with an encoder. It is driven by the L298N motor driver.

\begin{figure}
\centering
  \includegraphics[width=\linewidth]{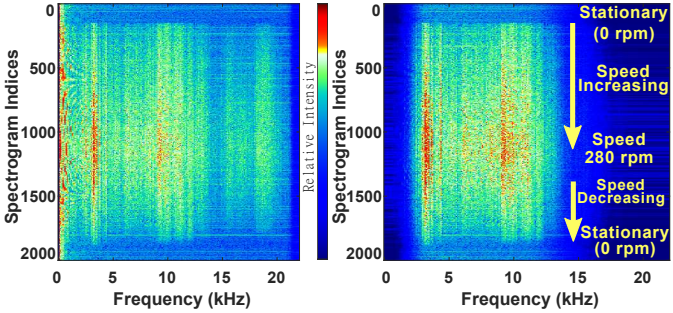}
  \caption{The spectrogram of the captured acoustic signals before (left) and after (right) filtering while the motor speeds up and slows down. }\label{fig_woot22_spec_prototype_wide}
\end{figure}

The attack system includes a microphone, a tweeter speaker, an audio amplifier, and a high-output-resolution sound card (e.g., Sound Blaster Z). 
The software of the attack system runs on a Linux desktop computer. It can also run on a Raspberry Pi with Linux installed.
The distance between the attack system and the victim system is 0.6 meter.
The attack system is not connected to the victim system through any wired or wireless control interfaces and communication channels.

\subsection{Experiments with Feedback}

\begin{figure}
\centering
  \includegraphics[width=0.958\linewidth]{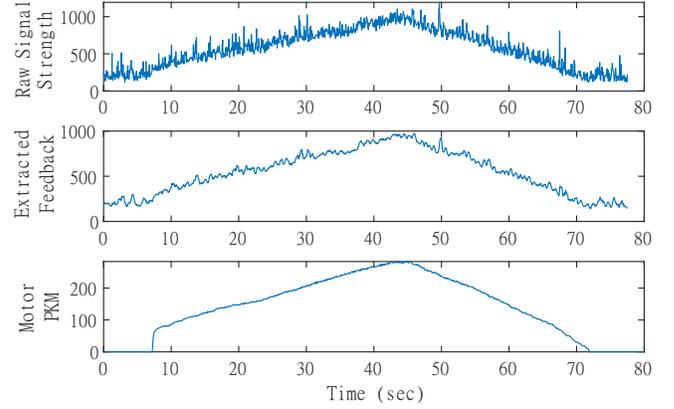}
  \vspace{-2.5mm}
  \caption{The extracted feedback from physical-domain signals (top and middle) is correlated with the motor speed of the victim (bottom).  }\label{fig_woot22_speed_dc_sound_wide}
\end{figure}

We analyze the feedback from the victim system, which uses a brushed, geared DC motor that is different from the brushless motor of self-balancing scooters. Our experimental results suggest that the motor speed can also be analyzed from its physical signal emanations (Fig. \ref{fig_woot22_spec_prototype_wide}). 
Furthermore, we control the motor with the micro-controller and record the actual speed using the motor encoder. We can observe that the extracted feedback is correlated with the motor speed (Fig. \ref{fig_woot22_speed_dc_sound_wide}). This correlation can be explained by our analysis of the motor speed and sound energy in Sec. \ref{woot22_subsubsec_feedback}.

We use the same methods as in experiments with the self-balancing scooter to derive the feedback.
In real-world attacks, the frequency range of the motor can be identified using a few seconds of audio recorded from the motor of the victim system (Fig. \ref{fig_woot22_spec_slowvsfast}). We also observe that using a subset of the motor sound components can derive similar results (Fig. \ref{fig_woot22_spec_prototype_narrow} and Fig. \ref{fig_woot22_speed_dc_sound_narrow}). For proof-of-concept attacks, we extract the feedback from acoustic signals only using a microphone. In the future, the feedback of the victim system can be analyzed more thoroughly by combining different kinds of measurements from cameras, radars, or other measuring devices.

\subsection{Experiments with Adversarial Process Control}
We study the attacks in the prototype testing system with program-controlled processes to manipulate the victim system.

Under the effect of resonant acoustic signals, the motor of the victim system accelerates and decelerates in an oscillating pattern. This is because the perceived heading angle of the system fluctuates under the perturbation of the injected signal.

\begin{figure}
\centering
  \includegraphics[width=\linewidth]{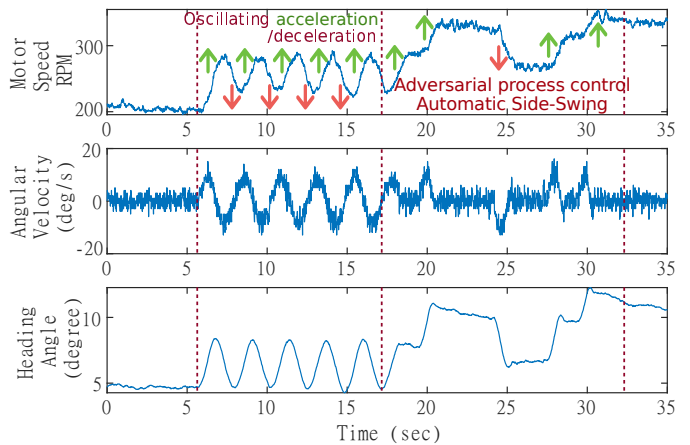}
  \caption{The speed and internal sensor measurements of the victim system in the process of adversarial control. The attack system cannot access these data. We can observe how the induced perturbations in the sensor (middle) affect the perception (bottom) and the actuation (top) of the victim control system in a continuous process. 
    }\label{woot22_processcontrol_victimdata1}
\end{figure}

Fig. \ref{woot22_processcontrol_victimdata1} shows the internal statuses of the victim system in this continuous process. We can observe that under the effect of acoustic resonant signals, the system's perceived heading angle fluctuates and falls back after each cycle. Under the interference, the victim system periodically accelerates and decelerates, and its speed fluctuates in an oscillating pattern.

\begin{figure}
\centering
  \includegraphics[width=\linewidth]{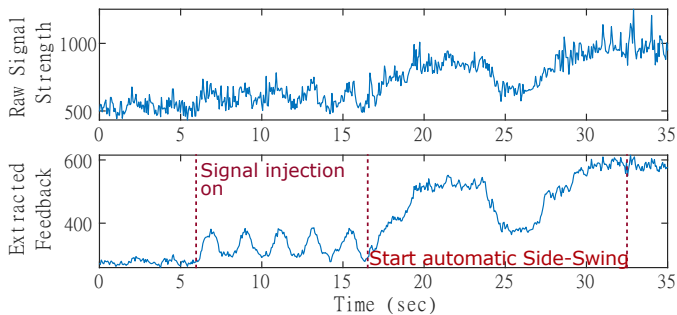}
  \caption{The system feedback is automatically extracted from physical-domain signals in the attack process. The induced oscillating movements (5.5-17 s) in the victim system perturbed by the signal injection can also be observed from the time series of the feedback.
  After the automatic Side-Swing process starts (from 17 s), the attack system automatically adjusts the attack signals to control the victim system. 
   }\label{woot22_processcontrol_attackdata1}
\end{figure}

The attack system leverages the feedback (Fig. \ref{woot22_processcontrol_attackdata1}) automatically extracted from physical-domain signals to guide the attack. It adjusts the attack signals in real time to control the motor speed without accessing the internal statuses of the victim system. After the automatic Side-Swing process starts, the attack system automatically performs amplitude adjusting of the attack signals within each cycle of the induced oscillation to selectively increase or decrease the perceived heading angle of the victim system in the real-time process.

Specifically, the attack system utilizes the recent feedback time series before starting the automatic Side-Swing attacks. It computes the period of the oscillation $p_0$ and an average value as a threshold $\sum\nolimits_{j=0}^{N} \frac{y_0[n-j]}{N}$ in the recent N samples of the time series (we use $N=100$ samples). It records the most recent time $T_0$ and direction when the feedback signal crosses the threshold. After the automatic Side-Swing attack starts, the system will adjust the amplitudes alternatively in real-time at an interval of half of the oscillation period ($\frac{p_0}{2}$) at specific times: $T_0 + k\frac{p_0}{2} - T_{offset} (k={1,2,3,...}) $, where $T_{offset}$ is the sum of a phase delay and a small real-time delay in the systems. The phase delay is $\frac{p_0}{4}$ between the oscillating injected signal in the gyroscope and the accumulated heading angle. Leveraging the time intervals, the system can adjust the amplitude of attack signals to inject signals in the target direction.

To demonstrate program-controlled processes in the adversarial control loop, we program the following procedure. The program in the attack system first controls the motor to speed up and maintain the speed. Then it controls the motor to slow down, and finally speed up and maintain the speed. The attack system automatically analyzes the extracted feedback and adjusts the attack signals to complete the procedure.

Fig. \ref{woot22_processcontrol_victimdata1} shows the process that the attack system gains control over the victim sensor-actuator system and performs the programmed procedure to apply changes to the motor speed in real time.
After the automatic process starts, the attack system selectively injects signals (angular velocity) into the target direction by swinging the amplitude to one specific side within each oscillation cycle. Instead of disrupting the perceived heading angle of the system, this controlled process leads to targeted changes in the heading angle and the motor speed.

The attack system does not access the internal statuses of the victim system. It utilizes the feedback extracted from physical-domain signals (Fig. \ref{woot22_processcontrol_attackdata1}) in the real-time attack process.

\section{Discussion}

\subsection{Limitation}

This paper investigates a general approach to overcoming the barriers that limit physical-level signal injection attacks on sensors from achieving system-level control in real-time, continuous processes. 
We do not focus on the underlying physics of how out-of-band signals are injected into the sensor systems. We do not address specific issues related to the signal power, attacker budget, and equipment.
Specific equipment such as long-range acoustic devices \cite{lrad,hypersonic} and radars could transmit the attack signals from a long distance. Signal power amplification methods had been investigated in prior works \cite{tu2018injected,xu2021inaudible,yan2019feasibility,roy2018inaudible}. We assume that the adversary is capable of inducing perturbations in the sensor signal through physical signals such as resonant acoustic or electromagnetic interference \cite{giechaskiel2019taxonomy,yan2020sok}.

We investigate how to utilize the system feedback automatically extracted from physical-domain signals to guide real-time attacks on sensor-actuator systems. We show that it is possible to construct an external adversarial control loop for process control in real-time sensor attacks.
However, the capability of the adversarial control loop is not as comprehensive and stable as the original internal control loop in the victim system. This is not only because of the lack of access to the internal statuses of the victim system but also due to the fact that our control mechanisms are constructed externally based on implicit analog channels instead of explicit interfaces that directly control the system. 
Additionally, we assume that the victim system has a feedback that can emit a measurable signal in the environment. Such signals can be acoustic, infrared, visual,  motion or electromagnetic signals, etc.  

Compared to attacks in white-box settings, our attack system does not have the access to the exact values of the motor speed, internal sensor reading and the perceived heading angle of the victim system. This means that we cannot fully control these statuses in an absolute manner. Instead, we control the victim system in a relative and coarse-grained manner. More capable and fine-grained control could be achieved with a more comprehensive external feedback extraction system. Such system could use cameras and lasers to measure the absolute speed of the motor. In the future, the feedback extraction can be conducted with a combination of different measuring devices to improve the robustness and capability of the attack system.

\subsection{Future Work}

We discuss possible future directions. First, by developing the external adversarial control loop and the automatic mechanisms, it is possible to construct a testbed that can facilitate automatically testing the security of sensing and actuation-based real-time control systems in a programmed, continuous process without connecting to the internal modules of embedded systems. Additionally, more robust feedback extraction methods can be developed in the future using a combination of measuring devices in attacks on complex systems and scenarios.
Furthermore, future works could extend our approach to other kinds of sensor attacks to gain control over real-time, continuous processes.

\section{Related Work}

From the perspective of control systems, sensors are abstracted resources that provide digital data streams to the system. Such abstraction determines that the systems do not necessarily understand the actual phenomena in the physical world, but rely on a restricted perspective based on the sensor data in the digital domain for functioning.

Researchers have utilized different kinds of physical signals such as electromagnetic, ultrasonic, and light signals in sensor attacks on smart voice assistants \cite{kune2013ghost,zhang2017dolphinattack,sugawara2020light,kasmi2015iemi,esteves2018remote,xu2021inaudible,yan2020surfingattack,ji2021capspeaker,wang2022ghosttalk}. These attacks explored the physical-level risks of exploiting sensors by transmitting determined signals (e.g., recorded voice) modulated in specific, out-of-band carriers to maliciously trigger an event in the victim system.

Our work investigates the threats of analog sensor attacks in real-time, continuous control processes. Many physical properties are controlled in a continuous process rather than an instantly triggered event. To control such processes, the attacker needs to continuously exert influence on the victim system to  apply changes to the physical environment in a real-time and controlled manner. 
However, the main issue of applying analog sensor attacks \cite{yan2020sok,giechaskiel2019taxonomy} 
to gain control over such processes
is that the exact time-varying values of injected signals are not fully predictable to external attackers. 

Our work investigates how to leverage the system feedback in an adversarial control loop to guide this real-time attack process without accessing the internal statuses of the victim system.

Prior works investigated manipulating sensing and actuation systems with physical-level signal injections \cite{selvaraj2018electromagnetic,son2015rocking,trippel2017walnut,bolton2018blue,tu2018injected}. For example, Selvaraj et al. studied electromagnetic induction attacks on servo motors to turn the motor position by synchronizing the injected voltage drops to the internal pulse-width modulation (PWM) signals of the victim system in a white-box setting \cite{selvaraj2018electromagnetic}.

Further, the actuation system of drones and smart devices can be disrupted by resonant acoustic attacks on the gyroscope \cite{son2015rocking,wang2017sonic}. 
An RC car can be commanded to move in a direction by saturating the smartphone accelerometer to spoof the remote controller application \cite{trippel2017walnut}. 
Moreover, Tu et al. demonstrated real-time attacks to control the kinematics and actuation of embedded systems based on inertial measurement units (IMUs) \cite{tu2018injected}. 
However, an external adversarial control loop over sensor-actuator systems was not investigated and constructed in prior works.
Different from prior attacks, our attack system leverages the time series of system feedback automatically extracted from physical-domain signals and program-controlled mechanisms to adjust the attack process in real time. Further, we study a general approach that can overcome the inherent limitations of analog sensor attacks to control real-time continuous processes.

Additionally, Shoukry et al. studied the risks of spoofing attacks on magnetic speed sensors of anti-lock braking systems to destabilize the vehicle \cite{shoukry2013non,shoukry2015pycra}. Researchers investigated the threats and defense methods of electromagnetic attacks on sensing and control systems \cite{tu2019trick,tu2021transduction,zhang2020detection,giechaskiel2019framework,barua2020hall}.
Recent works explored signal injection attacks on cameras \cite{kohler2021they,kohler2021signal,ji47poltergeist,yan2016can,man2020ghost} to spoof computer vision systems. Researchers also studied attacks on LiDAR systems to deceive the perception in autonomous vehicles \cite{cao2019adversarial,shin2017illusion,petit2015remote}.

\section{Conclusion}

This paper studied an adversarial control loop-based approach to manipulate sensor-actuator systems in continuous real-time attack processes. Different from the commonly-used internal control systems, the mechanisms of the adversarial control loop are constructed externally without connecting to the internal modules and statuses of the system.
The approach can overcome the real-time uncertainty that prevents analog sensor signal injections from achieving system-level adversarial control in continuous, real-time processes.

In our case study, we constructed a real-time attack system with different modules running in parallel. By automatically extracting and utilizing the time series of system feedback from physical-domain signals, the programs in the external attack system can continuously adjust the attack signals to achieve targeted process control over the victim system.

\paragraph{Acknowledgement}
The authors thank the anonymous reviewers for their valuable comments that improved this paper.
This work is supported in part by the US NSF under grants OIA-1946231, CNS-2117785, and a gift from Facebook.

\ifCLASSOPTIONcaptionsoff
  \newpage
\fi

{
\bibliography{ts}{}
\bibliographystyle{IEEEtran} 
}

\newpage
\appendix

Anonymized demos of the proof-of-concept attacks are available at the following link \url{https://www.youtube.com/playlist?list=PL_l1Kb3yQ2-ZllwC31CqIG5dNzJTXObF_}.

\begin{figure}[h]
\centering
  \includegraphics[width=0.97\linewidth]{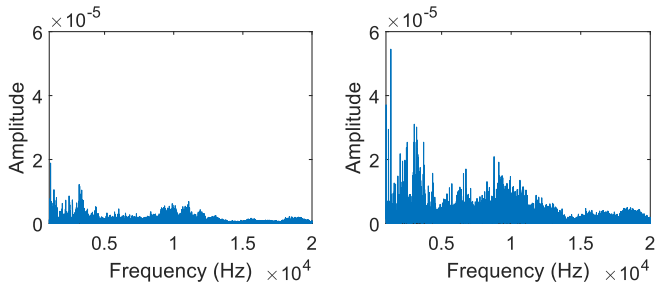}
  \caption{Signal spectrum of the motor acoustic emanations in a low speed (left)(about 100 rpm) and a high speed (right)(about 280 rpm). }\label{fig_woot22_spec_slowvsfast}
\end{figure}

\begin{figure}[h]
\centering
  \includegraphics[width=\linewidth]{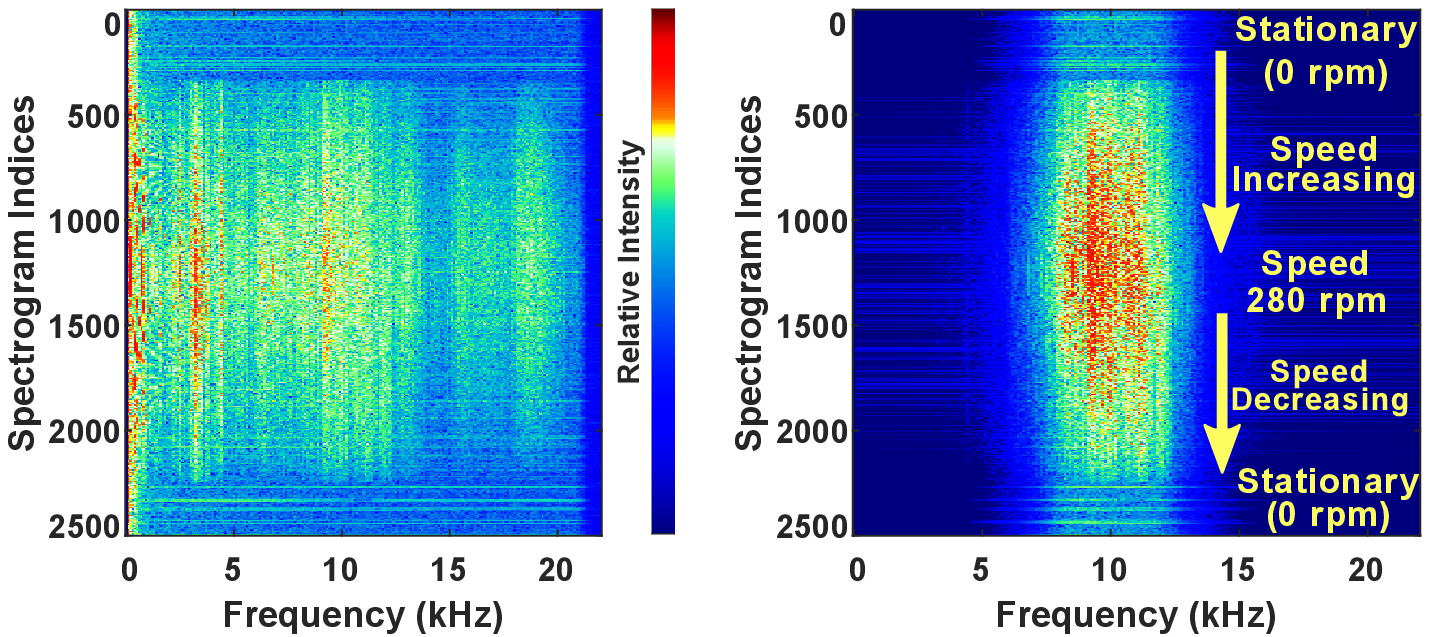}
  \caption{The spectrogram of the captured acoustic signals before (left) and after (right) filtering while the motor speeds up and slows down. }\label{fig_woot22_spec_prototype_narrow}
\end{figure}

\begin{figure}[h]
\centering
  \includegraphics[width=0.968\linewidth]{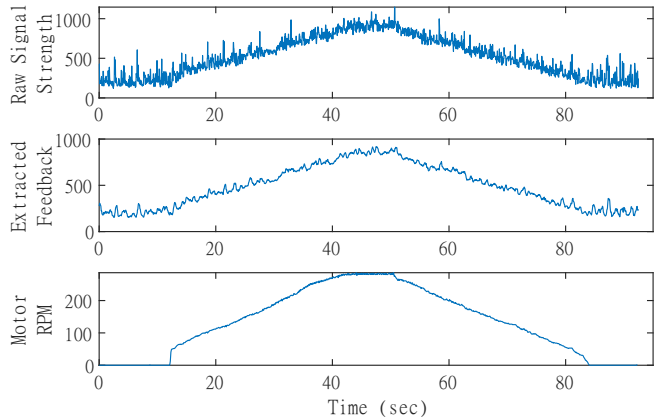}
  \caption{The extracted feedback from physical-domain signals is correlated with the motor speed using a subset of the motor sound components. }\label{fig_woot22_speed_dc_sound_narrow}
\end{figure}

\end{document}